\documentclass[aps,prl,reprint,superscriptaddress]{revtex4-1}
\usepackage{amsmath,braket}
\usepackage{graphicx}
\usepackage{color}
\usepackage{hyperref}
\usepackage[separate-uncertainty=true,multi-part-units=single]{siunitx}

\definecolor{darkgreen}{rgb}{0.0,0.4,0.0}
\DeclareSIUnit{\mu}{\micro\meter}
\DeclareSIUnit{\unit}{\relax}

\begin{document}


\title{Coherent and robust high-fidelity generation of a biexciton in a quantum dot by rapid adiabatic passage}



\author{Timo Kaldewey}
\affiliation{Department of Physics, University of Basel, Klingelbergstrasse 82, CH-4056 Basel, Switzerland}

\author{Sebastian L\"{u}ker}
\affiliation{Institut f\"{u}r Festk\"{o}rpertheorie, Universit\"{a}t M\"{u}nster, Wilhelm-Klemm-Strasse 10, D-48149 M\"{u}nster, Germany}

\author{Andreas V.\ Kuhlmann}
\affiliation{Department of Physics, University of Basel, Klingelbergstrasse 82, CH-4056 Basel, Switzerland}
\affiliation{IBM Research-Zurich, S\"{a}umerstrasse 4, 8803 R\"{u}schlikon, Switzerland}

\author{Sascha R.\ Valentin}
\affiliation{Lehrstuhl f\"{u}r Angewandte Festk\"{o}rperphysik, Ruhr-Universit\"{a}t Bochum, D-44780 Bochum, Germany}

\author{Arne Ludwig}
\affiliation{Lehrstuhl f\"{u}r Angewandte Festk\"{o}rperphysik, Ruhr-Universit\"{a}t Bochum, D-44780 Bochum, Germany}

\author{Andreas D.\ Wieck}
\affiliation{Lehrstuhl f\"{u}r Angewandte Festk\"{o}rperphysik, Ruhr-Universit\"{a}t Bochum, D-44780 Bochum, Germany}

\author{Doris E. Reiter}
\affiliation{Institut f\"{u}r Festk\"{o}rpertheorie, Universit\"{a}t M\"{u}nster, Wilhelm-Klemm-Strasse 10, D-48149 M\"{u}nster, Germany}

\author{Tilmann Kuhn}
\affiliation{Institut f\"{u}r Festk\"{o}rpertheorie, Universit\"{a}t M\"{u}nster, Wilhelm-Klemm-Strasse 10, D-48149 M\"{u}nster, Germany}

\author{Richard J.\ Warburton}
\affiliation{Department of Physics, University of Basel, Klingelbergstrasse 82, CH-4056 Basel, Switzerland}



\date{\today}

\begin{abstract}
A biexciton in a semiconductor quantum dot is a source of polarization-entangled photons with high potential for implementation in scalable systems. Several approaches for non-resonant, resonant and quasi-resonant biexciton preparation exist, but all have their own disadvantages, for instance low fidelity, timing jitter, incoherence or sensitivity to experimental parameters. We demonstrate a coherent and robust technique to generate a biexciton in an InGaAs quantum dot with a fidelity close to one. The main concept is the application of rapid adiabatic passage to the ground state-exciton-biexciton system. We reinforce our experimental results with simulations which include a microscopic coupling to phonons.
\end{abstract}

\pacs{}

\maketitle 

Entangled photon pairs are a powerful resource, especially for quantum teleportation and quantum key distribution protocols. Spontaneous parametric down-conversion in non-linear optics is a source of entangled photon pairs \cite{Kwiat1995}, but success is not guaranteed -- the emission is a probabilistic process -- and the error rate is high. In contrast, semiconductor quantum dots (QDs) are bright, on-demand sources of both single photons \cite{He2013} and entangled photon pairs and hence have enormous potential in quantum computing and quantum cryptography \cite{Gisin2002}.

\begin{figure}[t]{}
	\centering
	\includegraphics[width=1\columnwidth]{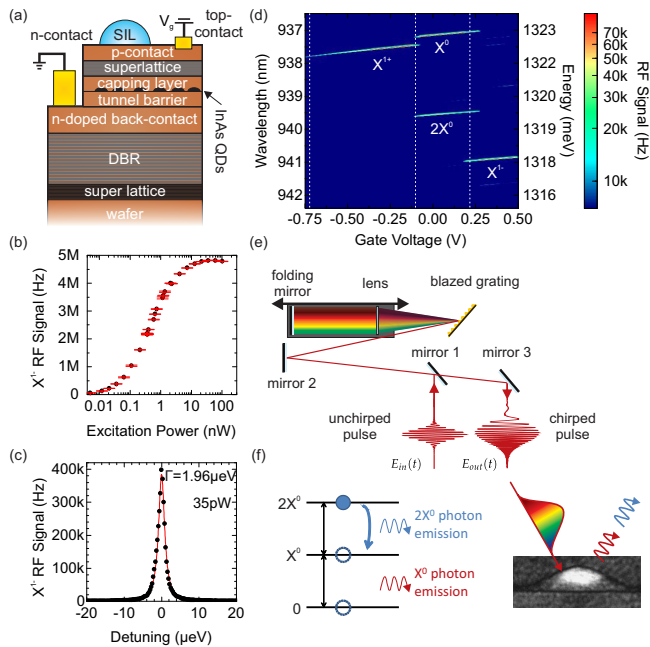}
	\caption{
	\textbf{Resonance fluorescence (RF) on a single quantum dot excited with a narrowband continuous wave laser and with broadband laser pulses.}
	\textbf{(a)} Structure of the n-i-p diode with epitaxial gate.
	\textbf{(b)-(c)} RF with narrowband excitation. \textbf{(b)} $\mathrm{X}^{1-}$ RF signal versus excitation power.
	\textbf{(c)} RF versus laser detuning at an excitation power of \SI{35}{\pico\watt} (points). A Lorentzian fit (red solid line) determined the FWHM linewidth, \SI{1.96}{\micro\electronvolt}.
		\textbf{(d)} RF with broadband pulsed excitation as a function of gate voltage. The laser polarization was linear, the chirp positive and the center frequency resonant with the neutral exciton $\mathrm{X}^{0}$.
	\textbf{(e)} Scheme of a folded $4f$ pulse-shaper controlling the chirp introduced into an ultra-short, transform-limited laser pulse.
	\textbf{(f)} Generation of a biexciton with broadband, chirped excitation. Scheme of the two-photon cascade (left) after chirped excitation of a QD (right).
	}
	\label{fig:fig1}
\end{figure}

A biexciton in a QD is the starting point for a two photon cascade: when perfectly prepared, biexciton decay leads to the subsequent emission of two photons, Fig.\ \ref{fig:fig1}(f). In a QD without a significant fine structure splitting (FSS), the two photons are polarization-entangled \cite{Muller2014}. The majority of InGaAs QDs show a FSS due to a reduced symmetry \cite{Gammon1996,Li2000_JAP,Juska2013}. However, sophisticated techniques were developed to compensate for the FSS with strain \cite{Trotta2014}, electric \cite{Kowalik2005} or magnetic fields \cite{Bayer2002,Stevenson2006} and with special growth conditions \cite{Juska2013}.

Several approaches for biexciton preparation have been proposed \cite{Glassl2013,Debnath2013,Gawarecki2012,Hui2008} and demonstrated \cite{Brunner1994,Stufler2006,Jayakumar2013,Muller2014,Gotoh2013,Bounouar2015}. Resonant two-photon schemes involving Rabi rotations \cite{Stufler2006, Jayakumar2013,Muller2014} are sensitive to fluctuations in both laser power and QD optical frequency. They are likely to suffer from an imperfect biexciton preparation resulting in undesired exciton photons unrelated to the cascade process. 

A more robust scheme using phonon-assisted excitation was reported by several groups recently \cite{Ardelt2014,Gotoh2013,Quilter2015,Bounouar2015,Jayakumar2013}. An impressively high biexciton occupation of up to 95\% was demonstrated using this quasi-resonant scheme \cite{Bounouar2015}. But the strength here is also a weakness. The scheme relies on the coupling to the phonon bath in the semiconductor environment: it is an inherently incoherent process. Also, a dependence on relaxation processes in the state preparation results in a timing jitter. In some cases, charge carrier relaxation times can reach values of up to a nanosecond  \cite{Reithmaier2014}.

We present here a coherent technique to create a biexciton with high probability, low jitter and weak dependence on the excitation and system parameters. The technique is based on rapid adiabatic passage (RAP). RAP allows the robust creation of an exciton \cite{Simon2011,Wu2011,Mathew2014} via a process requiring two-levels. RAP is applied here to the ground state-exciton-biexciton system, $\ket{0} - \ket{\mathrm{X}^{0}} -\ket{\mathrm{2X}^{0}}$, a three-level system \cite{Hui2008,Glassl2013}, and allows biexciton creation without significant exciton creation. In the implementation here, we use the full bandwidth of ultra-short \SI{130}{\femto\second} pulses allowing us to access spectrally both the ground state-exciton and exciton-biexction transitions within one laser pulse. The broad bandwidth pulses have the advantage over the narrow bandwidth pulses suggested in \cite{Glassl2013} of enhanced robustness owing to stronger avoided crossings and a decoupling of phonons even for negative chirped pulses. We describe this process theoretically and demonstrate excellent agreement with the experimental results. Moreover, we analyze the influence of phonons on the preparation protocol. For RAP-based exciton creation, the influence of phonons depends sensitively on the sign of the chirp \cite{Luker2012,Mathew2014,glassl2011lon}. We find also in RAP-based biexciton creation that we can choose the chirp such that the phonons are unimportant at low temperature.

We study self-assembled InGaAs QDs at a temperature of \SI{4.2}{\kelvin}. The QDs are grown by molecular beam epitaxy and embedded in an n-i-p or n-i-Schottky structure, Fig.\ \ref{fig:fig1}(a) (with more details in the supplementary information (SI) \cite{SI}). The bias voltage allows control over the QD charge via Coulomb blockade; and within a charging plateau, control over the optical resonance frequency via the DC Stark shift, Fig.\ \ref{fig:fig1}(d). The biexciton binding energy is positive and in the order of a few meV, a typical feature for InGaAs QDs. On driving the optical resonance with a narrowband continuous wave laser and detecting the resonance fluorescence (RF), we find that QDs in both samples have linewidths below \SI{2}{\micro\electronvolt}, Fig.\ \ref{fig:fig1}(c), close to the transform limit \cite{Kuhlmann2013_NatPhys}. Above saturation on a QD in the n-i-p sample, Fig.\ \ref{fig:fig1}(a), we detect a RF single photon count-rate of \SI{5}{\mega\hertz}, Fig.\ \ref{fig:fig1}(b). In the RAP experiments, we excite single QDs with the full bandwidth of \SI{130}{\femto\second} pulses with a center wavelength of around \SI{940}{\nano\meter} and linear polarization. The spectral full-width-at-half-maximum (FWHM) of the pulses is $\Delta \lambda=\SI{10}{\nano\meter}$. This allows us to address the exciton and biexciton optical transitions (but not the transitions involving higher shells in the QD) with just one laser pulse. The transform-limited pulses (repetition rate of \SI{76}{\mega\hertz}) from a passively mode-locked laser were manipulated in a folded $4f$ pulse-shaper \cite{Martinez1987}, Fig.\ \ref{fig:fig1}(e), in order to introduce chirp \cite{SI}. We control precisely the sign and magnitude of the chirp \cite{SI}. The FWHM of the pulse duration in intensity is stretched up to $\Delta t=\SI{15}{\pico\second}$ covering chirp coefficients up to $|\alpha|=\SI{0.70}{\pico\second\squared}$ For more details about the set-up, we refer to the SI \cite{SI}.

\begin{figure}[t]{}
	\centering
	\includegraphics[width=1\columnwidth]{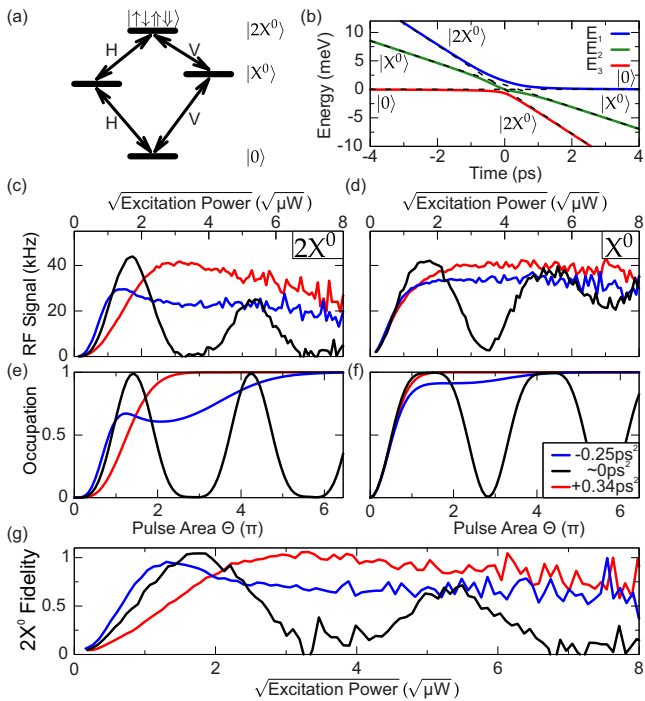}
	\caption{\textbf{Biexciton generation on QD1.} 
	\textbf{(a)} Energy level scheme. 
	\textbf{(b)} Instantaneous eigenenergies, the dressed states. 
	\textbf{(c)} $\mathrm{2X}^{0}$ emission and 
	\textbf{(d)} $\mathrm{X}^{0}$ emission as a function of the time-averaged square-root excitation power for different chirp parameters. 
	\textbf{(e)} and 
	\textbf{(f)} Simulation of the biexciton occupation and the sum of exciton and biexciton occupations as a function of the pulse area.
	\textbf{(g)} Ratio of the experimental data: $\mathrm{2X}^{0}/\mathrm{X}^{0}$.
	}
	\label{fig:fig2}
\end{figure}

The electronic structure in QDs is in general complex. For example in II-VI semiconductor colloidal QDs the electronic structure of multi exciton complexes was studied \cite{Kambhampati2012} and several biexciton states were identified \cite{Sewall2009}. However, in the III-V semiconductor QDs studied here, only the s-shells are populated such that the system can be described by four states, namely the ground state $\ket{0}$, the two bright exciton states $\ket{\mathrm{X}^{0}_{H/V}}$ and a single biexciton $\ket{\mathrm{2X}^{0}}$, Fig.~\ref{fig:fig2}(a). The p-shells are energetically far removed even when using the full bandwidth of the ultrafast laser pulses. The energy of the biexciton state $\ket{\mathrm{2X}^{0}}$ is $\hbar\omega_{2\mathrm{X}^{0}} = 2\hbar\omega_{\mathrm{X}^{0}} - \Delta_B$ where $\hbar\omega_{\mathrm{X}^0}$ is the energy of the neutral exciton and $\Delta_B$ the biexciton binding energy. These energies are determined experimentally from the emission spectrum. Considering only one linear polarization, H or V, and assuming that the laser pulses are much faster than the fine structure-induced quantum beat, a three-level system with only one exciton state, $\ket{\mathrm{X}^{0}}$, is sufficient. 

The exciton and biexciton state are coupled to longitudinal acoustic (LA) phonons in the pure dephasing regime via the deformation potential \cite{SI}. We use standard GaAs parameters and take the excitation parameters from the experiment leaving the QD size as the only fitting parameter. This model has been successfully used to describe the phonon influence on Rabi rotations \cite{Reiter2014} and on the population inversion via RAP \cite{Luker2012,Glassl2013}. For the occupations of the exciton and biexciton states as well as for the coherences between all involved states a set of equations of motion is derived within a well-established fourth-order correlation expansion method \cite{krugel2005the}. The equations are then solved numerically.

The goal of our work is high fidelity biexciton generation using RAP. Ideally, a vanishing intermediate occupation of the exciton is required. Less critical is a residual occupation of the ground state: this reduces only the rate of entangled photon emission but not the entanglement. A figure of merit is the ratio of biexciton photons to exciton photons. The ideal case is a ratio of one where every exciton photon is part of the two photon cascade on biexciton decay, Fig.\ \ref{fig:fig1}(f). In determining a biexciton creation fidelity we assume that biexciton and exciton photons are created and detected with equal probability. This is likely to be a very good assumption. The biexciton could in principle decay by an Auger process but in practice the probability of an Auger process is very small (probability of 0.23\% \cite{Kurzmann2016}); and a re-excitation process during the cascade decay is also very unlikely on account of the hierarchy of times (the pulse duration (up to \SI{15}{\pico\second}) is much shorter than the radiative lifetime (\SI{1}{\nano\second}) which is much shorter than the time separating successive pulses (\SI{13}{\nano\second})). Therefore, we define here the fidelity as the ratio of detected photons from the transition $\left|\mathrm{2X}^0\right>\rightarrow\left|\mathrm{X}^0\right>$ ($2\mathrm{X}^0$ photon) and from the transition $\left|\mathrm{X}^0\right>\rightarrow\left|0\right>$ ($\mathrm{X}^0$ photons).

Fig.\ \ref{fig:fig2} shows data from a single QD in the n-i-Schottky device. The laser pulses are centered at the two-photon biexciton resonance at \SI{944.3}{\nano\meter}. The $\mathrm{2X}^0$ and $\mathrm{X}^0$ emission intensities and their ratio are shown as a function of the square-root of the excitation power in Fig.\ \ref{fig:fig2}(c), (d) and (g), respectively. For close-to-zero chirp (black curves in Fig.\ \ref{fig:fig2}), Rabi rotations are observed. At the first maximum, the $\mathrm{X}^0$ and $\mathrm{2X}^0$ emission intensities are equal to within an error of 5\%, then both curves go down nearly to zero. At the second maximum, the $\mathrm{2X}^0$ emission reduces to 75\% of the $\mathrm{2X}^0$ emission. For higher excitation power, we observe mainly $\mathrm{X}^0$ emission with little emission from the $\mathrm{2X}^0$. At the first maximum it is clear that ultrafast pulses with close-to-zero chirp enable high fidelity preparation of the biexciton by a Rabi rotation. However, this $\pi$-pulse excitation is very sensitive to variations in the detuning or excitation power. We turn to RAP which potentially offers a more robust scheme. 

To create the biexciton using RAP, we first concentrate on positive chirp. Introducing a chirp of +\SI{0.34}{\pico\second\squared} stretches the pulse to an intensity FWHM of \SI{8}{\pico\second}. In the experiment, the $\mathrm{2X}^0$ emission rises more slowly with pulse power than in the Rabi rotation experiment (red curve in Fig.\ \ref{fig:fig2}) but then reaches a very broad maximum. Both $\mathrm{2X}^0$ and $\mathrm{X}^0$ signals reach a common maximum where the signals correspond closely to the maximum achieved in the Rabi rotation experiment. These are the main signatures of RAP. In terms of $\mathrm{2X}^0$ preparation, the ratio of $\mathrm{2X}^0$ to $\mathrm{X}^0$ emission reaches the ideal case of 100\% with an error of 5\%, proving that we can achieve a high fidelity biexciton preparation with RAP, robust against power and detuning fluctuations.

We now focus on the response of the system to a negative chirp of \SI{-0.25}{\pico\second\squared} (pulse duration of \SI{6}{\pico\second}, blue curves in Fig.\ \ref{fig:fig2}). In this case the $\mathrm{2X}^0$ signal rises at smaller pulse areas than for positive chirp and in fact at even smaller pulse areas than the Rabi rotations. The maximum of the $\mathrm{2X}^0$ signal is considerably less than the signal following a $\pi$ Rabi pulse. The maximum $\mathrm{X}^0$ signal is likewise reduced. The highest $\mathrm{2X}^0$ to $\mathrm{X}^0$ ratio is 94\% decreasing rapidly at high pulse areas. A negative chirp works less well at $\mathrm{2X}^0$  creation than a positive chirp. However, at higher excitation powers the difference between positive and negative chirp becomes less and less pronounced. 

An analysis of the calculated instantaneous eigenenergies, the dressed states Fig.\ \ref{fig:fig2}(b), allows a qualitative understanding of the physics. Starting in the ground state $\ket{\mathrm{0}}$, the system evolves along the lower (red) branch for positive chirp. Provided that the pulse area is above the RAP threshold, the system remains in the lower branch at the $\ket{\mathrm{0}}$ and $\ket{\mathrm{2X}^{0}}$ avoided crossing such that the system ends up in the biexciton state $\ket{\mathrm{2X}^{0}}$. Although there is no direct coupling between the $\ket{0}$ and $\ket{\mathrm{2X}^{0}}$ states, the avoided crossing arises as $\ket{0}$ and $\ket{\mathrm{2X}^{0}}$ have a common coupling to the $\ket{\mathrm{X}^{0}}$ state. Phonons can interrupt the adiabatic transfer by causing a jump from one branch to the other one \cite{Glassl2013}. For positive chirp, the system starts in the lowest energy dressed state and only phonon absorption is possible. At \SI{4.2}{\kelvin} however, phonons at the relevant energy scale are frozen out and the probability for absorption is small. Hence the process with positive chirp is barely affected by phonons at low temperature. Conversely, for negative chirp the system starts out in the upper-most branch (blue curve in Fig.\ \ref{fig:fig2}(b) with time running from ``right" to ``left"); the system can now jump to lower branches by phonon emission, a process which is possible even at low temperature. This leads to a significant probability of exciton and ground state population, reducing the fidelity of the RAP process. All these observations correspond well to the experimental data. 

The theory provides a quantitative account of three-level RAP in the presence of phonon coupling. We calculate the $\mathrm{2X}^0$ and $\mathrm{X}^0$ occupations as a function of the pulse area: the $\mathrm{2X}^0$ signal is proportional to the $\mathrm{2X}^0$ occupation; the $\mathrm{X}^0$ signal is proportional to the sum of the $\mathrm{2X}^0$ and $\mathrm{X}^0$ occupations. The simulation results are also shown in Fig.\ \ref{fig:fig2} and reproduce the main features of the experimental data extremely well, notably the exact form of the Rabi oscillations for close-to-zero chirp (a broad minimum at pulse area $\sim 3 \pi$ in the $\mathrm{2X}^0$ signal, a broad maximum at $\sim 1.5\pi$ in the $\mathrm{X}^0$ signal); the ``delayed" (``accelerated") rise of the $\mathrm{2X}^0$ signal for positive (negative) chirp; the relative signal strengths; and the better RAP performance for positive chirp. In the theory, the Rabi rotations are not influenced significantly by phonons because the Rabi dynamics are too fast for the phonons to follow. The theoretical results show clearly that $\mathrm{2X}^0$ generation with RAP is strongly (weakly) influenced by phonons for negative (positive) chirp for small to modest pulse areas. At the highest pulse areas, however, the theory predicts essentially perfect RAP independent of the sign of the chirp. The interpretation is that the splittings between the branches at the avoided crossings are energetically so large that they lie well above the energy range of the phonons which are efficiently coupled to the exciton system \cite{Reiter2014}. In other words, the wavelength of the phonons required for scattering between the dressed states becomes much smaller than the size of the QD at high pulse area. There is evidence that this reduction of the phonon efficiency is also seen in the experiment: the RAP signals for positive and negative chirp approach each other at the largest pulse areas. However, in addition the experimental signals exhibit a decay at high pulse areas which cannot be explained by the phonon coupling model. Instead, we tentatively attribute this decay to an occupation of higher energy levels by multi-photon absorption \cite{Patton2005}.

\begin{figure}[t]{}
	\centering
	\includegraphics[width=1\columnwidth]{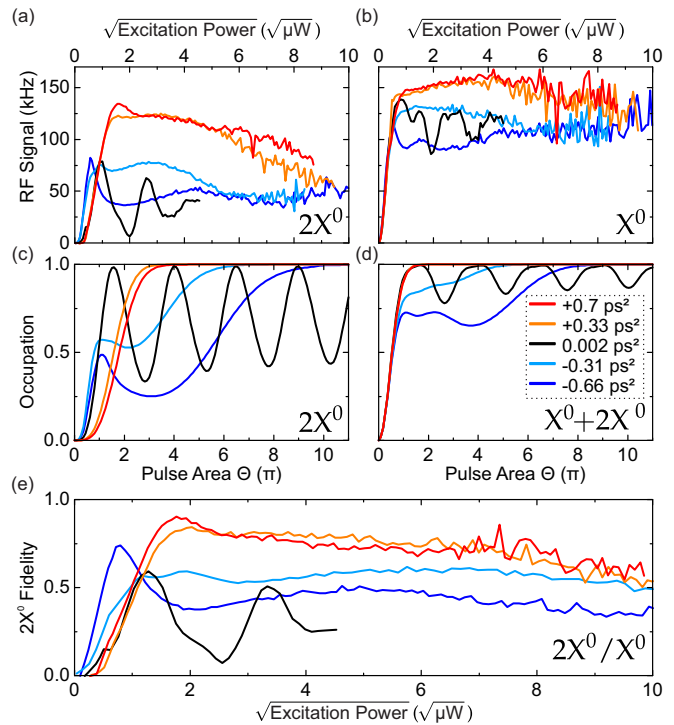}
	\caption{\textbf{Biexciton generation on QD2.} 
	\textbf{(a)} $\mathrm{2X}^{0}$ emission and 
	\textbf{(b)} $\mathrm{X}^{0}$ emission as a function of the time-averaged square-root excitation power for different chirp parameters. 
	\textbf{(c)} and 
	\textbf{(d)} Simulation of the biexciton occupation and the sum of exciton and biexciton occupations as a function of the pulse area.
	\textbf{(e)} Ratio of the experimental data: $\mathrm{2X}^{0}/\mathrm{X}^{0}$.
	}
	\label{fig:fig3}
\end{figure}

\begin{figure}[t]{}
	\centering
	\includegraphics[width=1\columnwidth]{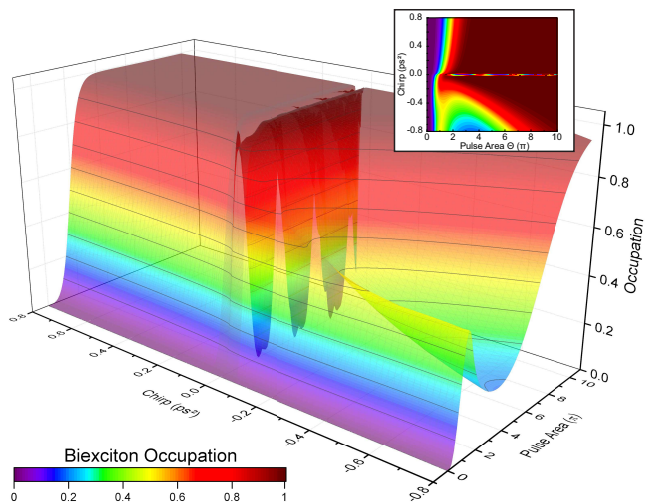}
	\caption{\textbf{Calculated biexciton generation on QD2.} 
	Simulation of the biexciton occupation as a function of the chirp parameter and the pulse area. The inset shows the top-view with the same color-scale.
	}
	\label{fig:fig4}
\end{figure}

To stress-test biexciton generation with RAP, we probe a second QD with much larger biexciton binding energy. RAP is more difficult in this case: the pulse area required for efficient $\mathrm{2X}^0$ generation increases, potentially entering the regime in which the additional decay process is active. The second QD, embedded in the n-i-p sample Fig.\ \ref{fig:fig1}(a), with biexciton resonance at wavelength \SI{938.2}{\nano\meter}, has a biexciton binding energy of \SI{3.4}{\milli\electronvolt} (the first QD has a binding energy of \SI{1.5}{\milli\electronvolt}). The $\mathrm{2X}^0$ emission intensity of the RAP experiment is shown in Fig.\ \ref{fig:fig3}. Despite the large biexciton binding energy, the ratio of the $\mathrm{2X}^0$ to $\mathrm{X}^0$ signals reaches 90\%, Fig.\ \ref{fig:fig3}(e). The general behavior is the same as for the first QD: a Rabi rotation-like behavior for minimal chirp and a distinct difference between positive and negative chirp. Again, apart from the decay at high pulse areas, the theory accounts extremely well for the experimental data. Notably, we find that for stronger negative chirp values, the phonons also become more effective resulting in a broader and deeper minimum of biexciton occupation which shifts to higher pulse areas, Fig.~\ref{fig:fig4}. Further, the RAP onset occurs at smaller pulse areas for negative chirp than for positive chirp, Fig.~\ref{fig:fig4}, a feature of the experimental data in Fig.~\ref{fig:fig3}(a).

In conclusion, we have demonstrated the coherent generation of a biexciton in a semiconductor quantum dot using a single chirped laser pulse. The state preparation has a very high fidelity over a broad range of excitation powers. The sign of the chirp is important: the scheme is robust with respect to phonon scattering at low temperature for positive chirp. A negative chirp results in damping due to phonon scattering. Theoretical calculations including a microscopic coupling to phonons reproduce all the experimental features apart from a damping in the experiment at high pulse areas.

\begin{acknowledgments}
We acknowledge financial support from EU FP7 ITN S$^{3}$NANO, NCCR QSIT and SNF project 200020\_156637. AL and ADW acknowledge gratefully support from DFH/UFA CDFA05-06, DFG TRR160 and BMBF Q.com-H 16KIS0109.
\end{acknowledgments}


%

\end{document}



\title{Coherent and robust high-fidelity generation of a biexciton in a quantum dot by rapid adiabatic passage: Supplementary Information}



\author{Timo Kaldewey}
\affiliation{Department of Physics, University of Basel, Klingelbergstrasse 82, CH-4056 Basel, Switzerland}

\author{Sebastian L\"{u}ker}
\affiliation{Institut f\"{u}r Festk\"{o}rpertheorie, Universit\"{a}t M\"{u}nster, Wilhelm-Klemm-Strasse 10, D-48149 M\"{u}nster, Germany}

\author{Andreas V.\ Kuhlmann}
\affiliation{Department of Physics, University of Basel, Klingelbergstrasse 82, CH-4056 Basel, Switzerland}
\affiliation{IBM Research-Zurich, S\"{a}umerstrasse 4, CH-8803 R\"{u}schlikon, Switzerland}

\author{Sascha R.\ Valentin}
\affiliation{Lehrstuhl f\"{u}r Angewandte Festk\"{o}rperphysik, Ruhr-Universit\"{a}t Bochum, D-44780 Bochum, Germany}

\author{Arne Ludwig}
\affiliation{Lehrstuhl f\"{u}r Angewandte Festk\"{o}rperphysik, Ruhr-Universit\"{a}t Bochum, D-44780 Bochum, Germany}

\author{Andreas D.\ Wieck}
\affiliation{Lehrstuhl f\"{u}r Angewandte Festk\"{o}rperphysik, Ruhr-Universit\"{a}t Bochum, D-44780 Bochum, Germany}

\author{Doris E. Reiter}
\affiliation{Institut f\"{u}r Festk\"{o}rpertheorie, Universit\"{a}t M\"{u}nster, Wilhelm-Klemm-Strasse 10, D-48149 M\"{u}nster, Germany}

\author{Tilmann Kuhn}
\affiliation{Institut f\"{u}r Festk\"{o}rpertheorie, Universit\"{a}t M\"{u}nster, Wilhelm-Klemm-Strasse 10, D-48149 M\"{u}nster, Germany}

\author{Richard J.\ Warburton}
\affiliation{Department of Physics, University of Basel, Klingelbergstrasse 82, CH-4056 Basel, Switzerland}



\date{\today}


\pacs{}

\maketitle 


\beginsupplement

\section{Experimental Methods}

\subsection{The quantum dot samples}
We studied in this work self-assembled InGaAs quantum dots (QD) in two different heterostructures. In both cases the QDs are embedded in a GaAs diode grown by molecular beam epitaxy. One diode is an n-i-Schottky structure \cite{Warburton2000, Drexler1994}, the other an n-i-p structure \cite{Prechtel2016}, Fig.\ \ref{fig:sfig1}(a) and (b), respectively. The n-i-Schottky diode uses an n-doped back-contact and a semi-transparent metallic top-contact (Ti/Au 3 nm/7 nm). The n-i-p sample on the other hand has an epitaxial top gate made from carbon-doped GaAs. In addition, the n-i-p diode features a distributed Bragg reflector (DBR). The layer structures are given in Table \ref{tab:niSchottky} and Table \ref{tab:nip} for the n-i-Schottky and n-i-p diode, respectively. In the n-i-p diode, the final layer, \SI{44}{\nano\meter} undoped GaAs, places the p-doped layer around a node-position of the standing electromagnetic wave in the n-i-p diode. 

In the experiment, the n- and p-doped layers are contacted independently. Selective etching of the capping allows the buried p-layer to be contacted. Access to the n-layer is provided by wet etching of a mesa structure. In both devices the n-contact is grounded and a gate voltage $V_g$ is applied to the top gate. The gate voltage allows control over the QD charge via Coulomb blockade and excitonic resonances can be shifted by the DC Stark effect \cite{Warburton2000,Dalgarno2008,Warburton2013}. The performance of single quantum dots (QDs) in both diode structures is quite similar. The main advantage of the n-i-p diode over the n-i-Schottky device is a strongly enhanced photon extraction. This is achieved on the one hand by the DBR and on the other hand by the low-loss epitaxial gate.

\begin{table}[b]
	\begin{tabular}{llr}
\hline \hline 
	Function & Material & Thickness \\ \hline
	wafer & GaAs &\\
	buffer & GaAs & \SI{50}{\nano\meter}\\
	superlattice, 30 periods & AlAs, GaAs & \SI{120}{\nano\meter}\\
	spacer & GaAs & \SI{50}{\nano\meter}\\
	electron reservoir (back contact) & GaAs:Si & \SI{50}{\nano\meter}\\
	tunnel barrier & GaAs & \SI{25}{\nano\meter}\\
	InGaAs QDs & InAs & $\sim$1.6 ML\\
	capping layer & GaAs & \SI{150}{\nano\meter}\\
	blocking barrier, 68 periods & AlAs, GaAs & \SI{272}{\nano\meter}\\
	capping layer & GaAs & \SI{10}{\nano\meter}\\
	\hline \hline
	\end{tabular}
	\caption{Layers of the heterostructure, n-i-Schottky diode. The quantum dots (QD) are formed by depositing 1.6 monolayer (ML) of InAs.}
	\label{tab:niSchottky}
\end{table}

\begin{table}[t]
	\begin{tabular}{llr}
\hline \hline 
	Function & Material & Thickness \\ \hline
	wafer & GaAs &\\
	buffer & GaAs & \SI{50}{\nano\meter}\\
	superlattice, 18 periods & GaAs, AlAs a \SI{2}{\nano\meter} & \SI{72}{\nano\meter}\\
	DBR, 16 periods & GaAs, AlAs & \SI{2400}{\nano\meter}\\
	spacer & GaAs & \SI{57.3}{\nano\meter}\\
	electron reservoir (back contact) & GaAs:Si & \SI{50}{\nano\meter}\\
	tunnel barrier & GaAs & \SI{30}{\nano\meter}\\
	InGaAs QDs & InAs & $\sim$1.6 ML\\
	capping layer & GaAs & \SI{153}{\nano\meter}\\
	blocking barrier, 46 periods & AlAs, GaAs & \SI{184}{\nano\meter}\\
	p-doped top contact & GaAs:C & \SI{30.5}{\nano\meter}\\
	undoped spacer & GaAs & \SI{1}{\nano\meter}\\
	etch stop & AlAs & \SI{2}{\nano\meter}\\
	capping layer & GaAs & \SI{44}{\nano\meter}\\
	\hline \hline
	\end{tabular}
	\caption{Layers of the heterostructure, n-i-p diode.}
	\label{tab:nip}
\end{table}

\subsection{Resonance fluorescence with continuous wave excitation}
\begin{figure}[t]{}
	\centering
	\includegraphics[scale=1]{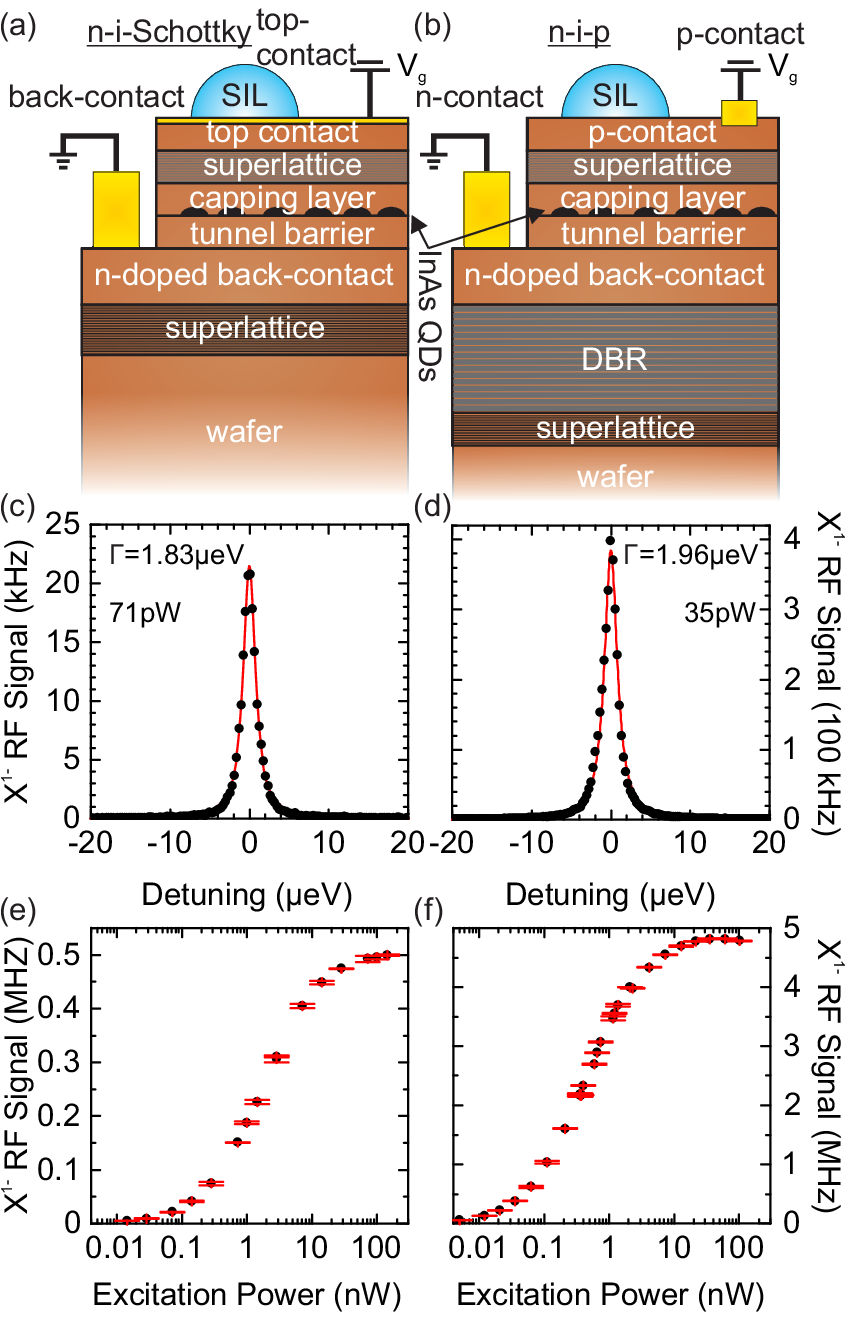}
	\caption{
	\textbf{Resonance fluorescence (RF) spectroscopy on a quantum dot (QD) with narrowband continuous wave excitation.}
	\textbf{(a)} n-i-Schottky, and 
	\textbf{(b)} n-i-p layer structure.
	\textbf{(c)-(f)} Comparison of the RF performance with narrowband continuous wave excitation of a single QD in the n-i-Schottky (left) and n-i-p structure (right). 
	\textbf{(c)-(d)} Example spectra showing the RF signal as function of laser detuning at an excitation power of \SI{71}{\pico\watt} (c), \SI{35}{\pico\watt} (d). A Lorentzian fit (red solid line) to the black data points shows a FWHM of \SI{1.83}{\micro\electronvolt} (c), \SI{1.96}{\micro\electronvolt} (d). The linewidths are similar in both devices.
	\textbf{(e)-(f)} RF signal as a function of excitation power for the n-i-Schottky device (e), and n-i-p device (f). The saturation count rate is a factor of ten higher in the n-i-p sample.
	}
	\label{fig:sfig1}
\end{figure}

In resonance fluorescence (RF) spectroscopy we investigated the quantum dot characteristics with a narrow band (linewidth in sub \si{\pico\meter} range) excitation laser.
The scattering induced by resonant excitation is detected. Back-reflected laser light is suppressed with a dark-field technique leading to an extinction ratio of $10^7$:1 \cite{Kuhlmann2013_NatPhys,Kuhlmann2013_RSI}. Typical $\mathrm{X}^{1-}$ spectra at low excitation power from single QDs in the two samples are depicted in Fig.\ \ref{fig:sfig1}(c) and (d). The linewidths are very similar, around \SI{2}{\micro\electronvolt}, approximately 2.5 times larger than the transform limit \cite{Kuhlmann2013_NatPhys}. A signal:background ratio of more than 1,000:1 is achieved with linearly polarized excitation. A more distinct difference between QDs in the two devices appears in the saturation behavior, Fig.\ \ref{fig:sfig1}(e) and (f): we detected single photons with a count-rate of \SI{5}{\mega\hertz} from the n-i-p sample, about 10 times more than from the n-i-Schottky sample. The high photon extraction efficiency of the n-i-p device, around 10\%, results from two improvements in the sample design as discussed above: the n-i-p type features a DBR and also a transparent epitaxial gate.

\subsection{Resonance fluorescence with pulsed excitation}

\begin{figure}[t]{}
	\centering
	\includegraphics[scale=1]{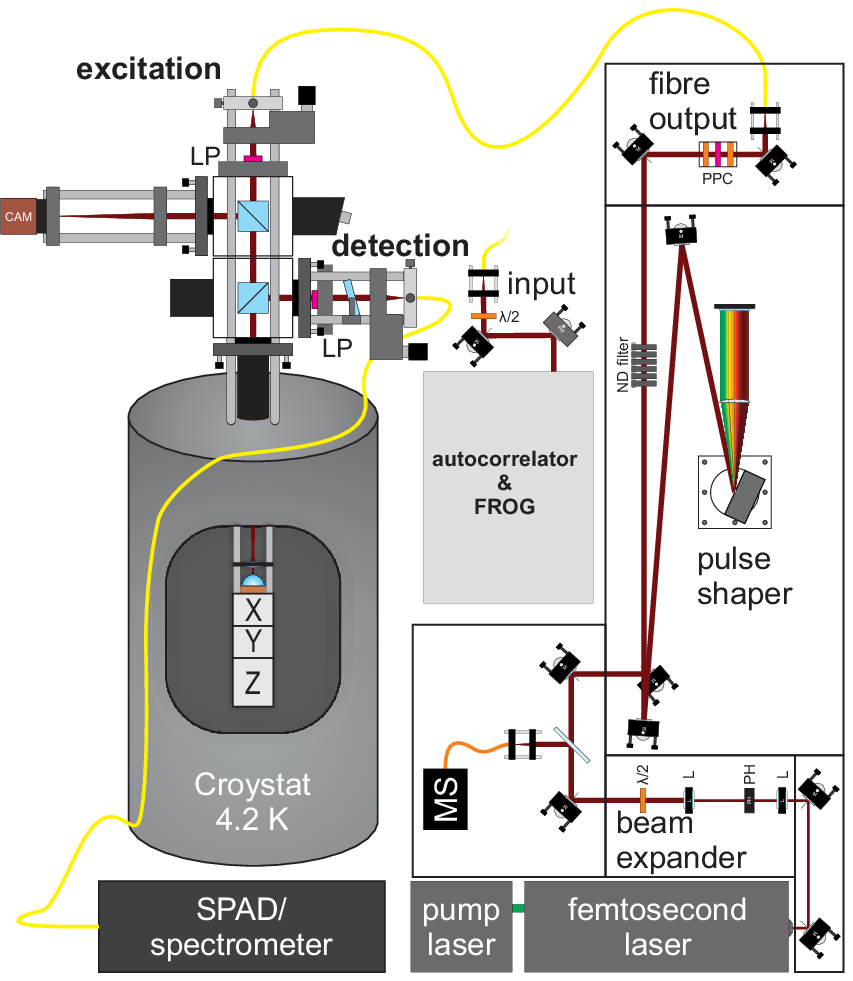}
	\caption{
	\textbf{Scheme of the complete set-up.} Ultra-short pulses were manipulated in a pulse-shaper, coupled into a single mode glass fiber and sent to a confocal microscope mounted on a bath cryostat. The pulses were characterized with a combination of autocorrelator and frequency-resolved-gating (FROG). The sample is held at a temperature of \SI{4.2}{\kelvin} in the bath cryostat on an xyz-positioner. A mini spectrometer (MS) measures the spectral profile of the pulses.
	}
	\label{fig:sfig2}
\end{figure}

With pulsed excitation, we use the full spectrum of \SI{130}{\femto\second} transform-limited ``ultra-fast" pulses, again exciting the QD resonantly. The spectral full-width-at-half-maximum (FWHM) of the pulse is $\Delta \lambda=\SI{10}{\nano\meter}$. With this bandwidth we address the ground state transition for the particular charge state set by the gate voltage (but not higher energy transitions). The RF response of the QD is detected with a grating spectrometer.

A scheme of the complete set-up is shown in Fig.\ \ref{fig:sfig2}. The ultra-fast pulses from a mode-locked femtosecond laser are expanded and sent into a compact pulse-shaper. The pulse-shaper retains all the spectral components and controls the amount of chirp as described below. From here the pulses pass through power control and polarization optics and are then coupled into a single mode optical fiber. The fiber transports the pulses to a confocal microscope. In the microscope, the excitation pulses pass through a linear polarizer, two beam-splitters and are sent to the objective at cryogenic temperature (4.2 K). The objective, an aspherical lens with a numerical aperture of 0.68, focuses the light onto the sample. The sample is held on a stack of piezo-steppers which allows a particular QD to be placed within the focal spot of the microscope. This process is aided by the in situ diagnostics provided by a camera image of the focus. Light scattered by the QD is collected and coupled into the detection fiber. The back-reflected laser light is suppressed by a second linear polarizer whose axis is orthogonal to the axis of the polarizer in the excitation stage. The RF signal is detected directly with a SPAD (continuous wave excitation) or with a spectrometer-CCD camera (pulsed excitation).

\begin{figure}[t]{}
	\centering
	\includegraphics[scale=1]{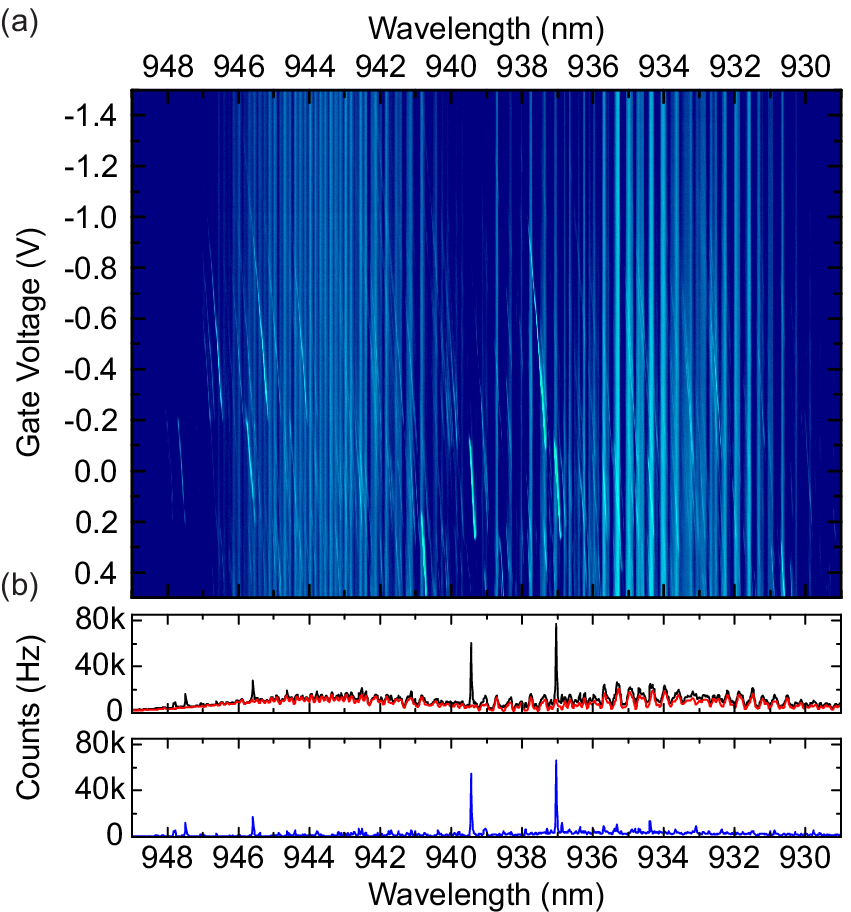}
	\caption{
	\textbf{Laser background subtraction resonance fluorescence spectroscopy using broadband, pulsed excitation.}
	\textbf{(a)} The RF signal as a function of gate voltage without background subtraction measured with an excitation power of \SI{\sim8}{\micro\watt} corresponding to a pulse area of around 3 $\pi$.
	\textbf{(b)} One spectrum from (a) at a gate voltage of \SI{-0.012}{\volt} is shown as black curve in the upper panel. Distinct peaks from the exciton and trion are visible together with a broad laser background. The reconstructed laser background (from all spectra in (a)) is shown in red. The blue curve in the lower panel shows the difference of the black and the red curve, the resulting signal after background subtraction.
	}
	\label{fig:sfig3}
\end{figure}

Detecting RF with pulsed excitation also depends on suppressing the reflected laser light with the polarization-based dark-field technique. However, owing to the wavelength dependence of the polarization optics, the laser suppression of a broadband pulsed laser is less effective than with the narrowband continuous wave laser: we reach an extinction ratio of typically $10^5$:1 with broadband excitation. We use in addition the spectral mismatch between the broadband laser and the narrowband RF to increase the extinction ratio, Fig.\ \ref{fig:sfig3}. The laser background falling on the CCD camera does not depend on the gate voltage, Fig.\ \ref{fig:sfig3}(a), such that we can use the gate voltage dependence of the QD emission to construct the laser background spectrum. The resulting residual laser background is shown in Fig.\ \ref{fig:sfig3}(b) in red. The spectral shape is explained by the Gaussian spectral shape of the excitation laser along with the quadratic function of laser suppression: the laser suppression is most effective at its alignment wavelength and then decays quadratically in wavelength superimposed by interference fringes. A typical spectrum measured at $V_g=\SI{-12}{\milli\volt}$ is shown in black. Sharp emission peaks from the exciton and biexciton appear on top of the laser background. The signal after background subtraction is shown in the lower plot in blue. Before background subtraction, the maximum signal to background ratio in a narrow spectral window around the emission line is 22:1.

\subsection{Pulse shaping: introducing chirp}\label{sec:pulseshaper}
Chirp is introduced into the transform-limited laser pulses by a compact, folded $4f$ pulse-shaper \cite{Martinez1987,Weiner1992}. The scheme is depicted in the main paper in Fig.\ 1. The unchirped pulse is diffracted by a high resolution, blazed grating (1800 grooves per mm) and then focused by a lens onto a mirror positioned in the focal plane. From here the light travels back under a small angle with respect to the diffraction plane allowing a spatial separation of incoming and outgoing pulses. The distance between grating and lens controls the chirp \cite{Martinez1987}. The lens and the mirror are mounted on a translation platform. Moving the platform with respect to the grating changes the chirp: if the distance matches the focal length of the lens the chirp is zero, a larger (smaller) distance leads to negative (positive) chirp. The temporal duration of the pulses exiting the laser is \SI{130}{\femto\second} (intensity FWHM); the temporal duration can be stretched up to \SI{15}{\pico\second} corresponding to a chirp in the range from \SI{-0.7}{\pico\second\squared} to \SI{+0.7}{\pico\second\squared}. 

We characterized the chirp with a combination of a FROG and autocorrelator. This allows us to reveal the presence of any high-order phase terms and prevent them by an optimal adjustment of the pulse-shaper. The chirp is measured both in the free space mode and also after the optical single mode fiber in order to determine the sign of the chirp and to compensate for the chirp introduced by the fiber itself.

\section{Theoretical Model}

\subsection{Hamiltonian}
The Hamiltonian is composed of four parts
\begin{align*}
H = H_{\textrm{c}} + H_{\textrm{c-l}} + H_{\textrm{ph}} + H_{\textrm{c-ph}},
\end{align*}
with $H_{\textrm{c}}$ denoting the electronic structure, $H_{\textrm{c-l}}$
the carrier-light coupling, $H_{\textrm{ph}}$ the free phonons and
$H_{\textrm{c-ph}} $ the carrier-phonon coupling.

We consider a three-level system consisting of the ground state $|0\rangle$,
the single exciton state $|\mathrm{X}^0\rangle$ and the biexciton state $|2\mathrm{X}^0\rangle$
yielding the Hamiltonian for the electronic system
\begin{align*}
H_{\textrm{c}} = \hbar\omega_{\mathrm{X}^{0}}|\mathrm{X}^{0}\rangle\langle \mathrm{X}^{0}|
+ \hbar\omega_{2\mathrm{X}^{0}}|2\mathrm{X}^{0}\rangle\langle 2\mathrm{X}^{0}|.
\end{align*}
Here, $\hbar\omega_{\mathrm{X}^0}$ is the energy of the neutral exciton and
$\hbar\omega_{2\mathrm{X}^{0}} = 2\hbar\omega_{\mathrm{X}^{0}} - \Delta_B$ is the energy of the
biexciton accounting for the biexciton binding energy $\Delta_B$; the ground
state energy has been set to zero. For the binding energy we take the
experimentally determined values (corrected for the polaron shift) $\Delta_B=1.4$~meV for QD1
and $\Delta_B=3.2$~meV for QD2.

We model the carrier-light interaction in the usual dipole and rotating wave
approximation resulting in the following Hamiltonian for the three-level
system:
\[
H_{\textrm{c-l}} =  \frac{1}{2}\hbar\Omega(t)\left(|\mathrm{X}^{0}\rangle\langle 0|
+ |2\mathrm{X}^{0}\rangle\langle \mathrm{X}^{0}| \right) + \textrm{h.c.},
\]
h.c. denoting the Hermitian conjugate. Here the instantaneous Rabi frequency
$\Omega(t)$ with $\hbar\Omega(t)=2\mathbf{M}\cdot \mathbf{E}(t)$ enters,
given by the positive frequency component of the electric field
$\mathbf{E}(t)$ and the dipole matrix element $\mathbf{M}$.

For the carrier-phonon interaction, we account for the deformation potential
coupling to longitudinal acoustic (LA) phonons which has been shown to be
the main source of decoherence in most typical QDs \cite{Reiter2014}. The
Hamiltonian of the free phonons is given by
\begin{align*}
H_{\textrm{ph}} = \hbar\sum_{\mathbf{q}}\omega_{\mathbf{q}}b^\dag_{\mathbf{q}}b_{\mathbf{q}},
\end{align*}
with the creation (annihilation) operators $b^\dag_{\mathbf{q}}$
($b_{\mathbf{q}}$) of a phonon with wave vector $\mathbf{q}$ and frequency
$\omega_{\mathbf{q}} = c_{\mathrm{LA}}|\mathbf{q}|$, $c_{LA}$ being the longitudinal
sound velocity.

The carrier-phonon interaction is described by the standard pure-dephasing
Hamiltonian. Assuming for the biexciton wave function a product of exciton
wave functions, the Hamiltonian reads
\begin{align*}
H_{\textrm{c-ph}} = \hbar\sum_{\mathbf{q},\nu}n_{\nu}|\nu\rangle\langle\nu|
\left(g_{\mathbf{q}}b_{\mathbf{q}} + g^*_{\mathbf{q}}b^\dag_{\mathbf{q}}\right),
\end{align*}
where the factor $n_{\nu}$ counts the number of the excitons which are
present in the state $|\nu\rangle$, i.e., $n_{\nu} = 0$ for
$|\nu\rangle=|0\rangle$, $n_{\nu} = 1$ for $|\nu\rangle=|\mathrm{X}^0\rangle$, and
$n_{\nu}=2$ for $|\nu\rangle=|2\mathrm{X}^{0}\rangle$.

The corresponding matrix elements for deformation potential coupling to
electrons and holes are
\[
	g_{\mathbf{q}}^{e/h}=\sqrt{\frac{q}{2V\rho\hbar c_{\mathrm{LA}}}}D_{e/h}F_{\mathbf{q}}^{e/h},
\]
with $V$ the normalization volume, $\rho$ the mass density of the crystal,
and $D_{e/h}$ the deformation potential coupling constants for
electrons/holes. We take GaAs parameters as listed in Table~\ref{tab:para}.
The exciton coupling matrix element is obtained from the electron and hole
coupling matrix elements by $g_{\mathbf{q}} = g_{\mathbf{q}}^{e}-
g_{\mathbf{q}}^{h}$.

The influence of the QD geometry is described in terms of the form factor
$F_{\mathbf{q}}^{e/h}$. We assume a spherical QD modeled by a harmonic
confinement potential yielding the form factors
\[
	F_{\mathbf{q}}^{e/h}=\exp\left(-q^2a_{e/h}^2/4\right),
\]
where $a_{e/h}$ are the electron/hole localization lengths. To fit the
experimental data we used $a_e=5$~nm and $a_h=2$~nm. We note that similar
values have been used in the theoretical modeling of phonon-assisted state
preparation schemes \cite{Quilter2015}.

\begin{table}[t]
\begin{tabular}{lcc}
\hline \hline 
Parameter & & Value \\ \hline
density & $\rho$& $5370\,$kg/m$^3$\\
sound velocity & $c_{\mathrm{LA}}$ & $5.1\,$nm/ps\\
electron deformation potential constant &$D_e$&$7$~eV\\
hole deformation potential constant &$D_h$&	$-3.5$~eV\\
\hline \hline 
\end{tabular}
\caption{\label{tab:para}Parameters used in the calculations.}
\end{table}

For the theoretical modeling of the light-induced dynamics we use the density
matrix formalism. The infinite hierarchy of phonon-assisted variables
appearing in the equations of motion is truncated using a fourth-order
correlation expansion. It has been shown that this method gives very
reliable results in a two-level system
\cite{Reiter2014,Luker2012,glassl2011lon}. Here we have extended this
approach to the three-level model. This extension is straightforward,
however the numerical complexity increases considerably. Based on this
formalism, we calculate the occupations of the biexciton
$\left<|2\mathrm{X}^{0}\rangle\langle 2\mathrm{X}^{0}|\right>$ and the neutral exciton
$\left<|\mathrm{X}^{0}\rangle\langle \mathrm{X}^{0}|\right>$ states. Note that also all
coherences between these states as well as one- and two-phonon assisted
density matrices are fully included in the calculation.

\subsection{Chirped laser pulses}
To model the optical excitation, we consider an electric field which
describes a linearly chirped Gaussian laser pulse. Such a chirped pulse is
generated from a transform-limited Gaussian laser pulse of the form
\begin{align*}
\Omega_0(t)=\frac{\Theta}{\sqrt{2\pi}\tau_0}\exp\left(\frac{t^2}{2\tau_0^2}\right)
\exp\left(-i\omega_L t\right)
\end{align*}
with pulse duration $\tau_0$, central laser frequency $\omega_L$, and pulse
area $\Theta$. After application of a chirp with coefficient $\alpha$ the
laser pulse is transformed to
\begin{align*}
\Omega(t)=\frac{\Theta}{\sqrt{2\pi\tau_0\tau}}\exp\left(\frac{t^2}{2\tau^2}\right)
\exp\left(-i\omega_L t - \frac{1}{2}a t^2\right).
\end{align*}
As a result of the chirp the pulse duration is increased to
$\tau=\sqrt{\frac{\alpha^2}{\tau_0^2}+\tau_0^2}$, while the central frequency
of the laser pulse changes with time according to the frequency chirp rate
$a=\frac{\alpha}{\alpha^2+\tau_0^4}$. The central laser frequency at the
pulse maximum is set to the arithmetic mean of the ground state-to-exciton
and exciton-to-biexciton transition frequencies, such that a two-photon process with this
frequency resonantly drives the biexciton.


%